\documentclass[review,12pt,authoryear]{elsarticle}
\usepackage{natbib}
\usepackage[utf8]{inputenc}
\usepackage{amsmath, amstext, amsthm, amsfonts, amssymb, mathrsfs}
\usepackage{lineno}
\usepackage{geometry}
\usepackage[pdfborder={0 0 0}]{hyperref}
\graphicspath{{figures/}} 
\usepackage{subfigure}
\renewcommand{\subfigtopskip}{0.1cm}
\renewcommand{\subfigcapskip}{0cm}
\usepackage{multicol}
\usepackage{float}
\usepackage{soul}
\usepackage[titletoc,title]{appendix}
\usepackage[section]{placeins} 
\usepackage[pdftex]{color}

\usepackage{booktabs}
\usepackage{nicefrac}
\usepackage{textcomp}

\begin{document}

\begin{frontmatter}

    \title{Invasive competition with Fokker-Planck diffusion and noise}

    \author[osna]{Michael Bengfort\corref{cor1}}
    \ead{michael.bengfort@uni-osnabrueck.de}
    \cortext[cor1]{Corresponding author}

    \author[goett]{Ivo Siekmann}
    \ead{ivo.siekmann@mathematik.uni-goettingen.de}

    \author[osna]{Horst Malchow}
    \ead{horst.malchow@uni-osnabrueck.de}
    
    \address[osna]{Institute of Environmental Systems Research, School of Mathematics\,/\,Computer Science,\\Osnabr\"uck University, Barbarastra\ss{}e 12, 49076 Osnabr\"uck, Germany}
    \address[goett]{Institute for Mathematical Stochastics, Georg-August-University of G\"ottingen,\\Goldschmidtstra\ss{}e 7, 37077 G\"ottingen, Germany}

    \begin{abstract}
Defeat and success of the competitive invasion of a populated area is described with a standard Lotka-Volterra competition model. The resident is adapted to the heterogeneous living conditions, i.e., its motion is modelled as space-dependent, so-called Fokker-Planck diffusion. The invader's diffusion is taken as neutral Fickian. Furthermore, it is studied how multiplicative environmental noise fosters or hinders the invasion.
    \end{abstract}


\end{frontmatter}



\section{Introduction}

Interactions and movements of populations in a heterogeneous and variable environment are often modelled with stochastic reaction-diffusion equations. Diffusive fluxes in ecology can differ due to specifics of the population's relationships and environmental heterogeneity. They might be neutral cf. eq.\,(\ref{eq:neutral}), attractive (\ref{eq:attract}) or repulsive (\ref{eq:repuls}), i.e., for $N$ populations
\begin{align}
\vec{j}_{in} &= - D_i(\vec{r},\mathbf{X})\, \vec{\nabla} X_i(\vec{r},t)\,,\label{eq:neutral}\\
\vec{j}_{ia} = - D_i^2(\vec{r},\mathbf{X})\, \vec{\nabla} \left[\dfrac{X_i(\vec{r},t)}{D_i(\vec{r},\mathbf{X})}\right] &= + X_i(\vec{r},t) \vec{\nabla} D_i(\vec{r},\mathbf{X}) - D_i(\vec{r},\mathbf{X})\, \vec{\nabla} X_i(\vec{r},t)\,, \label{eq:attract}\\
\vec{j}_{ir} = -\vec{\nabla} \left[D_i(\vec{r},\mathbf{X}) X_i(\vec{r},t)\right] &= - X_i(\vec{r},t) \vec{\nabla} D_i(\vec{r},\mathbf{X}) - D_i(\vec{r},\mathbf{X})\, \vec{\nabla} X_i(\vec{r},t)\,; \label{eq:repuls}\\
&~~~~~~~~~~~~~~~~~~~~~~~~~~~~~~~~~~~~~~~~~~~~~~i=1,2,\ldots,N. \nonumber
\end{align}

The usual notation is used: $\mathbf{X}(\vec{r},t)=\{X_i(\vec{r},t); ~i=1,2,\ldots,N\}$ is the vector of population densities at position $\vec{r}=\{x,y\}$ and time $t$ and $D_i(\vec{r},\mathbf{X})$ their possibly space- and density-dependent diffusion coefficient. The formulations (\ref{eq:neutral}--\ref{eq:repuls}) have been elaborated by \citeauthor{Ske:51} (\citeyear{Ske:51,Ske:73}, and nicely summarized by \citet{Oku:80}, see also \citet{Aro:85} and \citet{Mur:89}. In order to complete the list of ecodiffusive fluxes in heterogeneous media, one could add the flux in environmental potentials $U(\vec{r})$
\begin{equation}
\vec{j}_{ip} =  \vec{j}_{ik} + \gamma_i X_i(\vec{r},t)\vec{\nabla}\,U(\vec{r});~i=1,2,\ldots,N;\label{eq:potent}
\end{equation}
where $\gamma_i$ is called the coefficient of affinity of $X_i$ to the environment and index $k$ can be $n$, $a$ and $r$ respectively, i.e., one of the fluxes (\ref{eq:neutral}--\ref{eq:repuls}) can be applied. The minima of $U(\vec{r})$ correspond to preferable and, therefore, attracting habitats. The latter concept has been derived from the ideas of habitat value and environmental density \citep{Mor:71,Shi:78}.

The neutral diffusion is also called Fickian \citep{Fic:55} whereas the repulsive type is named after Fokker and Planck (\citeyear{Fok:14,Pla:17}).
For a certain density dependence of diffusion, the latter has been used for modelling the spatial segregation of populations \citep{Shi:79,Mim:80} as well as the formation of Turing patterns \citep{Mal:88c}.

In a recent publication \citep{Ben:16b}, the diffusivities have been assumed purely space-dependent. Spatial patterns may already occur without any interactions. For this setting, the spatially stationary solution has been derived. Furthermore, the speed of diffusive waves of a single logistically growing population has been analytically estimated, and conditions for the formation of spatio-temporal and Turing patterns in an excitable prey-predator system have been given.

Another recent publication \citep{Siek:16} has dealt with the control of invasion of a populated area by selective infection of the invader as well as by white and coloured noise-modulated environments the resident is adapted to but being unfavourable for the invading population.

The present work shall link the two latter approaches. The Lotka-Volterra textbook model of the competition of two populations is combined with space-dependent Fokker-Planck diffusion of the residents, Fickian diffusion of the invaders and environmental noise. It will be shown that the spatial heterogeneity modelled by Fokker-Planck diffusion but also the external noise can foster or hinder the invasion.


\section{The stochastic competition-diffusion model}

The dynamics of resident $X_1$ and invader $X_2$ is described by 
\begin{align}
\dfrac{\partial X_1}{\partial t}=&(1-X_1)X_1-c_{12}X_1X_2+d_1\nabla^2(X_1D^\ast(x,y))+g_1(X_1)\xi(\vec{r},t)\,,\label{eq:resi}\\
\dfrac{\partial X_2}{\partial t}=&(1-X_2)X_2-c_{21}X_1X_2+d_2\nabla^2 X_2 + g_2(X_2)\xi(\vec{r},t)\,.\label{eq:inva}
\end{align}

The space dependence of the resident's diffusivity is chosen as
\begin{equation}
\label{eq:diffu}
D^\ast(x,y)=D_0+\left\{\begin{array}{lc} a\left(\sin(\sqrt{x^2+y^2})\right)^m & \text{if }\sqrt{x^2+y^2}<3\pi\,,\\
			a\left(\sin(3\pi)\right)^m & \text{else\,.}\end{array}\right.
\end{equation}
This spatially varying diffusivity is meant to represent a simple fragmented landscape with a varying habitat quality for species $X_1$. The parameter $m$ is an even number witch controls the steepness of $D^\ast$.

For simplicity, just uncorrelated white noise $\xi(\vec{r},t)$ is applied here, i.e.,
\begin{equation}
\langle\xi(\vec{r},t)\rangle =0\,, \langle\xi(\vec{r}_1,t_1)\xi(\vec{r}_2,t_2)\rangle=\delta(\vec{r}_1-\vec{r}_2)\delta(t_1-t_2)\label{eq:noise1}
\end{equation}

with linearly density dependent noise intensities
\begin{equation}
g_i(X_i)=\omega_i X_i\,; i=1,2\,. \label{eq:noise2}
\end{equation}


\section{Numerical methods}
\subsection{Crank-Nicolson scheme for two dimensions with Fokker-Planck diffusion}
\label{sec:crank}

We split the Laplace operator into two parts.
First, we calculate the diffusion in one spatial dimension ($x$), second we do the same for the other spatial dimension ($y$).
\begin{equation}
\label{FPDiff}
\dfrac{\partial X}{\partial t} = \vec{\nabla}^2(XD) = \dfrac{\partial^2(XD)}{\partial x^2}+\dfrac{\partial^2(XD)}{\partial y^2},
\end{equation}

where $X$ is the population density and $D$ its spatially varying diffusion coefficient which can be written as
\begin{equation}
D(x,y)=d_1 D^\ast(x,y)
\end{equation}

with $d_1=const$ and $D^\ast(x,y)\neq 0~\forall~x,y$.
Now we formulate the Crank-Nicolson algorithm \citep{Cra:47} for one spatial dimension as follows
\begin{align}
\dfrac{X_k^{t+\Delta t}-X_k^t}{\Delta t}=\dfrac{d_1}{2\Delta x^2}&\left(X_{k+1}^{t+\Delta t}D^\ast_{k+1}-2X_k^{t+\Delta t}D^\ast_k+X_{k-1}^{t+\Delta t}D^\ast_{k-1}\right.\notag\\
&\left.+X_{k+1}^{t}D^\ast_{k+1}-2X_k^{t}D^\ast_k+X_{k-1}^{t}D^\ast_{k-1}\right).
\end{align}

Here $k\in(1,n)$ is the index of the spatial position of $X$, whereas $t$ is the time which varies with a discrete time step $\Delta t$.
With $\alpha=d_1\dfrac{\Delta t}{\Delta x^2}$ we can write this as a system of linear equations
\begin{equation}
\mathbf{A}\left(\vec{X}^{t+\Delta t}\vec{D}^\ast
\right)=\mathbf{B}\left(\vec{X}^t\vec{D}^\ast\right)
\end{equation}

where $\vec{X}$ and $\vec{D}^\ast$ are vectors of length $n$ including the values of $X_k$ and $D_k^\ast$ at each spatial position in one dimension $k\in(1,n)$. $\mathbf{A}$ and $\mathbf{B}$ are $n\times n$ tridiagonal matrices
\begin{align*}
\mathbf{A}&=\left(\begin{array}{ccccc}2\left(\dfrac{1}{D^\ast_1}+\alpha\right) & -\alpha & 0& \dots & 0\\
-\alpha & \ddots & -\alpha & 0 &\vdots\\
0 & -\alpha & \ddots & \ddots & 0\\
\vdots &\dots&\ddots & \ddots &-\alpha\\
0 &\dots & &-\alpha & 2\left(\dfrac{1}{D^\ast_n}+\alpha\right)\end{array}\right)\\
\text{and}&\\
\mathbf{B}&=\left(\begin{array}{ccccc}2\left(\dfrac{1}{D^\ast_1}-\alpha\right) & \alpha & 0& \dots & 0\\
\alpha & \ddots & \alpha & 0 &\vdots\\
0 & \alpha & \ddots & \ddots & 0\\
\vdots &\dots&\ddots & \ddots &\alpha\\
0 &\dots & &\alpha & 2\left(\dfrac{1}{D^\ast_n}-\alpha\right)\end{array}\right).
\end{align*}

This implicit scheme has been proven to be unconditionally stable for two spatial dimensions.

In order to implement zero-flux boundary conditions we have to add the term $-\alpha$ to the matrix components $\mathbf{A}_{11}$ and $\mathbf{A}_{nn}$, and the term $\alpha$ to the matrix components $\mathbf{B}_{11}$ and $\mathbf{B}_{nn}$.

To calculate the distribution of $X^t$ at time step $t+\Delta t$, we have to multiply the vector $\vec{X^t}$ with the spatially varying coefficient of diffusion $\vec{D}^\ast$ and solve the equation $\mathbf{A}\vec{Y}=\mathbf{B}\vec{X}$, where $\vec{X}$ is a input-vector (in our case $\vec{X^t}\cdot \vec{D}^\ast$) and $\vec{Y}$ is a output-vector. After that the components of the output-vector $\vec{Y}$ has to be divided with the corresponding components of the vector $\vec{D}^\ast$, which is temporally constant in order to get the distribution $X^{t+\Delta t}$. Once this scheme has been performed for each row in one spatial direction it has to be repeated for the other spatial dimension in every time step.


\subsection{Derivative-free Milstein method for interactions and noise}
\label{sec:milstein}

For numerical integration of the interaction and noise terms, the derivative-free Milstein method is used \citep{Mil:95,Klo:99}. The Milstein scheme reads for white noise (\ref{eq:noise1},\ref{eq:noise2}) with time step $\Delta t$ and in Stratonovich interpretation
\begin{align}
 X_i^{t+\Delta t} &= X_i^t + f_i(X_i^t) \Delta t + \omega_i X_i^t \Delta W_i + \dfrac{\omega_i}{2}\left[f_i(X_i^t) \sqrt{\Delta t} + \omega_i X_i^t \right] (\Delta W_i)^2\,,\\
 \text{with}&\nonumber\\
 \Delta W_i &= W_i^{t+\Delta t} - W_i^t \sim \sqrt{\Delta t} \, \mathcal{N}(0,1)\,.\nonumber
\end{align}
As usual, $\mathcal{N}(0,1)$ stands for the normal distribution with zero mean and unity variance. The required uniformly distributed random numbers are generated with the Mersenne Twister \citep{Mat:98b}, the normally distributed random numbers with the common Box-Muller algorithm \citep{Box:58}.


\section{Numerical simulations and results}

The following parameters have been applied:
$$D_0=1\,, ~m=8\,, ~c_{12}=c_{21}=1.2\,.$$
Because both species are described with the same parameter values, the difference in the coefficient of diffusion determines wether or not an invasion of species $X_2$ is successful in case of homogeneous $D$, i.e., $a=0$, cf. eq.\,(\ref{eq:diffu}).

\begin{figure}[!ht]
\center
\subfigure[$t=3900$\,; $\omega_1=\omega_2=0$]{\includegraphics[width=0.275\textwidth]{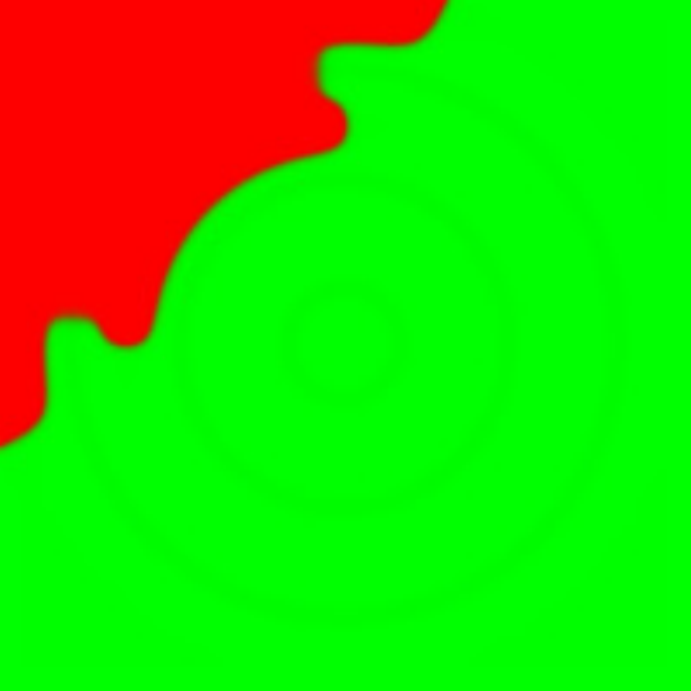}}\hfill
\subfigure[$t=3900$\,; $\omega_1=\omega_2=0.4$]{\includegraphics[width=0.275\textwidth]{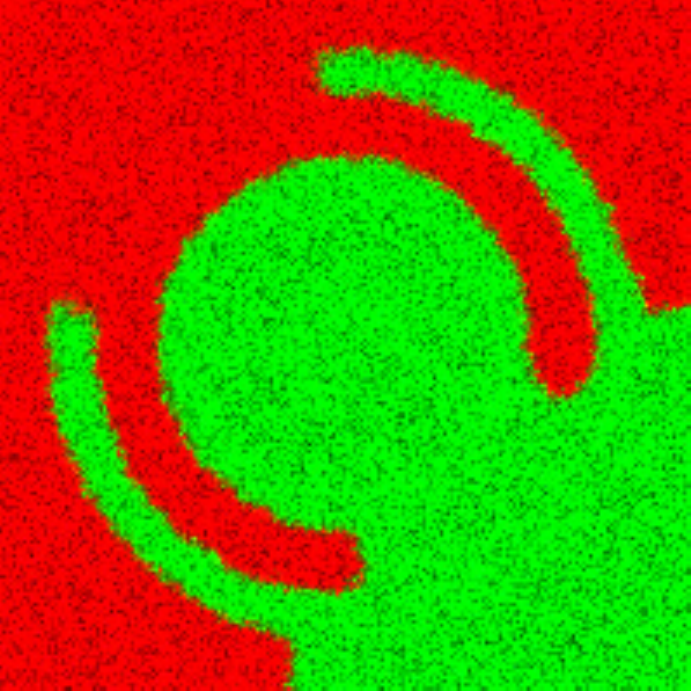}}\hfill
\subfigure[$t=600$\,; $\omega_1=\omega_2=0.6$]{\includegraphics[width=0.275\textwidth]{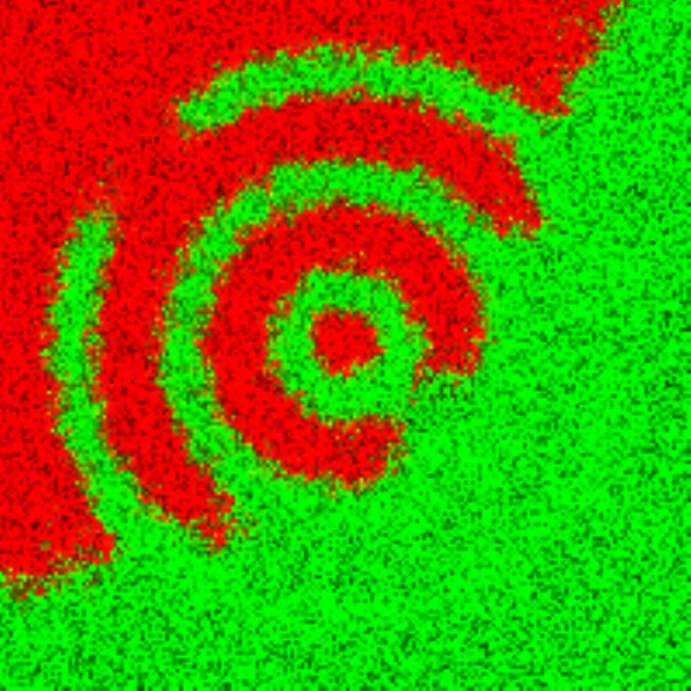}}
\caption{$d_1=5$, $d_2=25$, $a=9$: The density of the resident species is reduced in areas of large $D^\ast$. The invader successfully invades the space, where it has a larger coefficient of diffusion as the resident species. Density-dependent multiplicative noise accelerates the invasion in areas of small $D^\ast$. Areas with large $D^\ast$ act as a barrier for the invasion. Strong noise can break through these barriers and induces invasion of $X_2$ in the inner circles with small $D^\ast$.}
\label{fig:invas1}
\end{figure}

If the native species $X_1$ has a smaller coefficient of diffusion in certain areas of the domain, whereas in the other areas its coefficient of diffusion $d_1\cdot D^\ast$ is larger as the constant coefficient $d_2$, invasion is successful in those areas where the invader has the larger coefficient of diffusion (Fig.\,\ref{fig:invas1}).
Areas with a high diffusivity of the native species act as barrier for the invasion. This fits well earlier published results on diffusion-controlled competitive invasions \citep{Mal:11}.
In this scenario multiplicative density-dependent noise, as described in eq.(\ref{eq:noise1}) and (\ref{eq:noise2}), accelerates the speed of invasion (Fig.\,\ref{fig:invas1}b).
Strong noise can push the invader through the barriers of large resident diffusivity and induce invasions in other regions with low resident diffusivity.

\begin{figure}[!ht]
\center
\subfigure[$\omega_1=\omega_2=0$]{\includegraphics[width=0.275\textwidth]{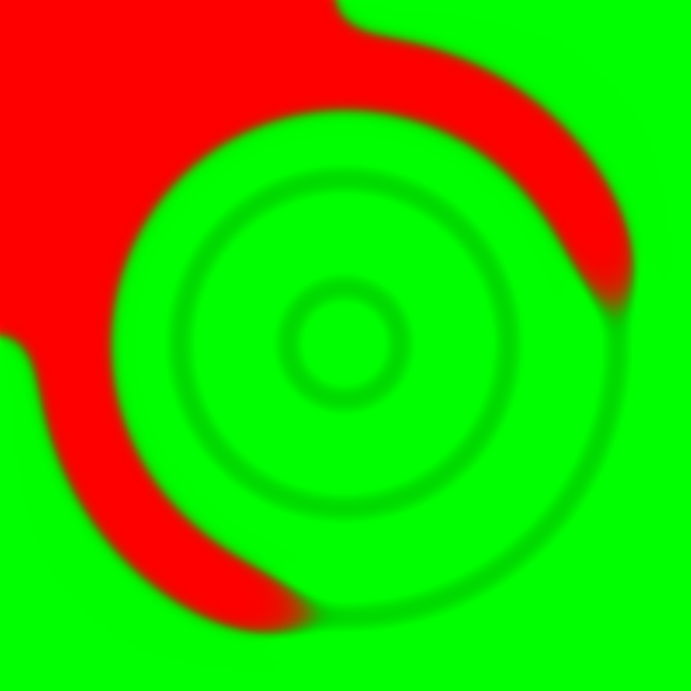}}\hspace*{1.25cm}
\subfigure[$\omega_1=\omega_2=0.25$]{\includegraphics[width=0.275\textwidth]{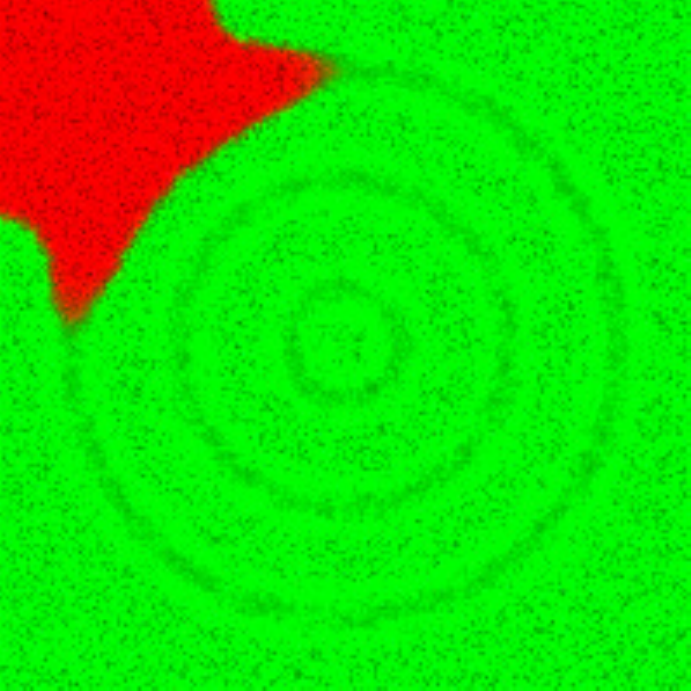}}
\caption{$t=4400$, $d_1=30$, $d_2=25$, $a=19$; large $\nabla^2 D(\vec{r})$: Due to the reduced resident concentration in areas of large $D^\ast$ invasion is possible even though the invader has a smaller coefficient of diffusion everywhere in the spatial domain. Noise reduces the invasion speed. Strong noise can invert the invasion.}
\label{fig:invas2}
\end{figure}

Because of the Fokker-Planck diffusion in eq.\,(\ref{eq:resi}), the spatial distribution of the resident species, $X_1$, develops proportional to $\nabla^2D^\ast(x,y)$, as described in \citet{Ben:16b}.
If this effect is strong enough, the reduced resident concentration in areas with high resident diffusivity enables an invasion of species $X_2$, even if the diffusivity of $X_1$ is larger than the diffusivity of $X_2$ everywhere in the domain (Fig.\,\ref{fig:invas2}). In this scenario, multiplicative density-dependent noise has a decelerating effect on the speed of invasion (Fig.\,\ref{fig:invas2}b).

\begin{figure}[!ht]
\center
\subfigure[Initial distribution]{\includegraphics[width=0.275\textwidth]{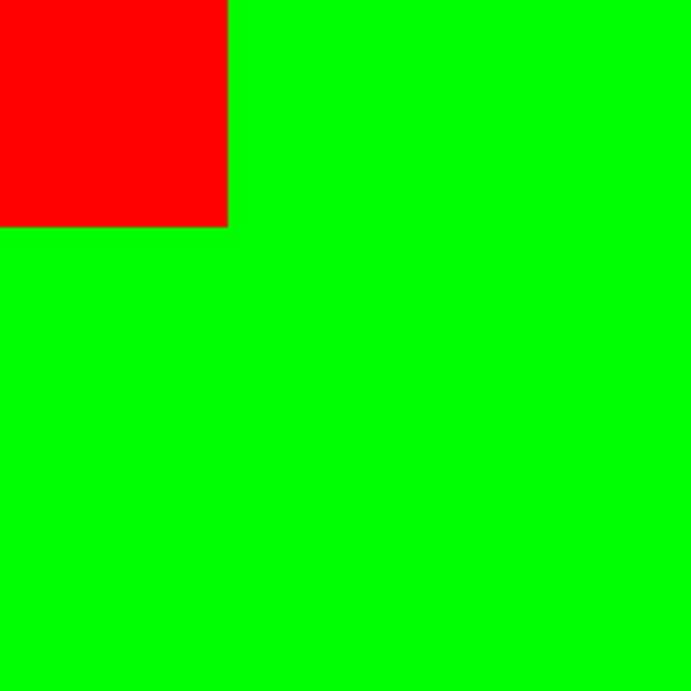}}\hfill
\subfigure[$D^\ast(x,y)$ for $a=9$]{\includegraphics[width=4.6cm]{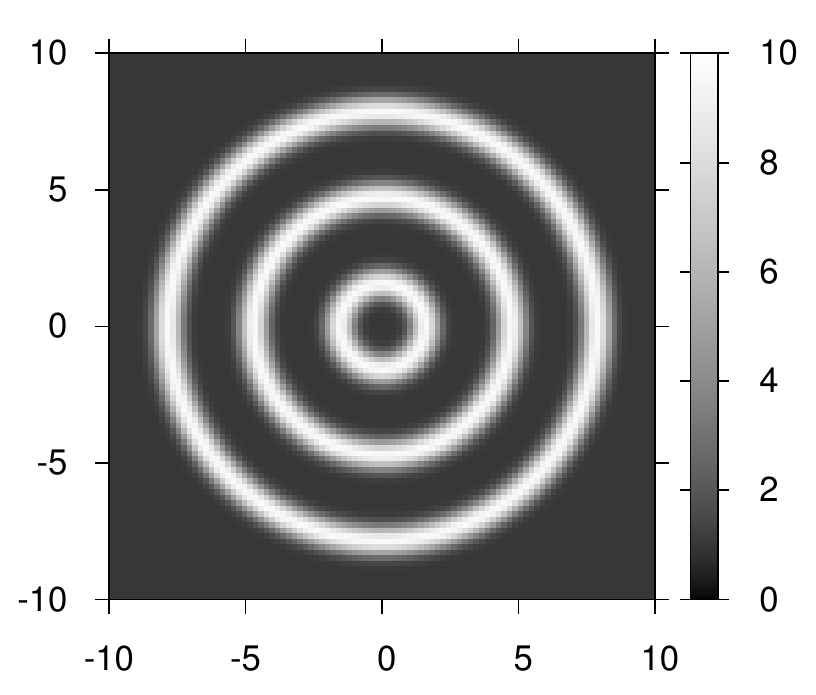}}\hfill
\subfigure[$\nabla^2D^\ast(x,y)$ for $a=9$]{\includegraphics[width=4.6cm]{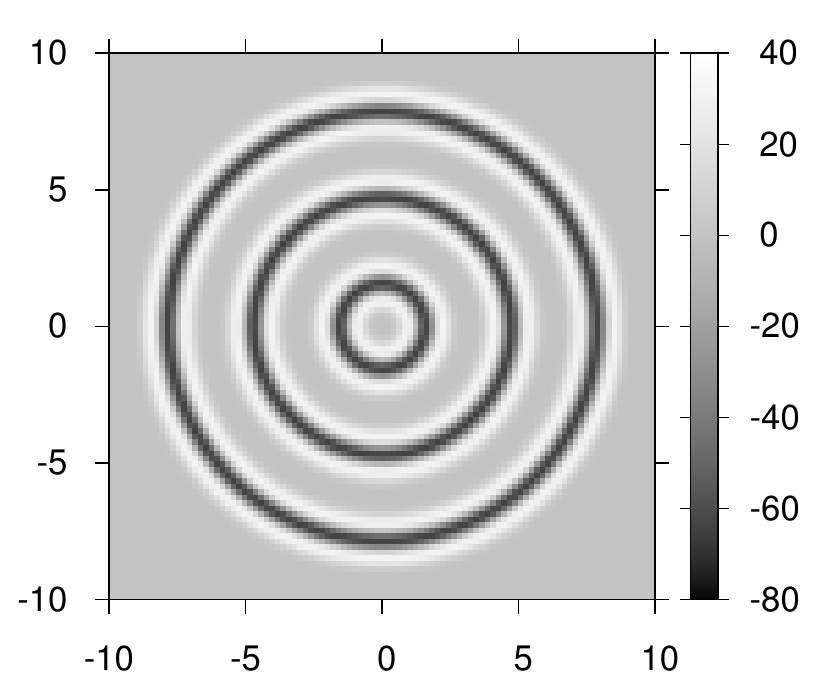}}
\caption{Initial settings for densities (green = resident, red = invader) and resident's diffusivity}
\label{fig:initial}
\end{figure}

\section{Conclusions}
It has been shown that a non-uniform diffusivity, i.e., Fokker-Planck diffusion, of a resident species in a spatially heterogeneous habitat can have different effects on the ability of a similar competing species to invade the habitat.

This Fokker-Planck type of modelling the movement of organisms generates patterns in the spatial population distribution which correspond to the spatial variation of the diffusion coefficient. If this effect is small, the competitor can invade the domain in areas where its (spatially constant) coefficient of diffusion is larger than that of the resident species.
This is not surprising because both species are described with equal parameters for growth and competition so that diffusivity determines the success of invasion if the size of the initial patch of the invading species exceeds the related critical patch size.
In a non-deterministic environment, where the populations are subject to stochastic fluctuations, the speed of invasion increases with increasing noise intensity. Strong noise can also induce invasions in areas which are perfectly protected against an invasion in the deterministic case. If the pattern forming effect of the Fokker-Planck diffusion is stronger, invasion is possible even though the coefficient of the invader is smaller than the one of the resident species everywhere in the domain. Contrary to the former example, noise has a negative effect on the success of invasion.
This is caused by the fact, that the density dependent noise counteract the pattern forming properties of the Fokker-Planck diffusion. The resident species benefits from the homogenising effect of the noise because it has a larger coefficient of diffusion than the invader.

In this paper, Gaussian noise in time and space was applied in order to model the variability of the environment. For future research it would be interesting to investigate the effect of spatially and/or temporally coloured noise in combination with the Fokker-Planck diffusion which generates patterns in the resident species with a certain wavelength.

Here, it was assumed that only the resident species favours certain areas in the domain and consequently move with a spatially varying speed and is therefore described with Fokker-Planck diffusion. One can also think of situations where the invader is described with a heterogeneous coefficient of diffusion as well. The areas favoured by the invading species can be the same as for the resident or independently distributed.

\section*{Acknowledgements}
The authors acknowledge the stimulating working and living conditions at the Mediterranean Institute of Oceanography during several visits of Aix-Marseille University. They appreciated the professional cooperation but also the sincere hospitality of Jean-Christophe Poggiale and his colleagues. Last but not least, H.M. is thankful for the perfect organization of MPDE'16. 


\bibliographystyle{model2-names}
\bibliography{refer}{}


\end{document}